\documentclass[12pt,english]{article}
\usepackage{avant}
\usepackage[T1]{fontenc}
\usepackage{geometry}
\usepackage{xcolor}
\usepackage{babel}
\usepackage{amsmath}
\usepackage{amsthm}
\usepackage{graphicx}
\usepackage[authoryear]{natbib}
\usepackage[unicode=true]
 {hyperref}

\makeatletter
\IfFileExists{candleshape.def}{%
 \input{candleshape.def}}{}
\IfFileExists{dropshape.def}{%
 \input{dropshape.def}}{}
\IfFileExists{TeXshape.def}{%
 \input{TeXshape.def}}{}
\IfFileExists{triangleshapes.def}{%
 \input{triangleshapes.def}}{}

\theoremstyle{plain}
    \ifx\thechapter\undefined
	    \newtheorem{thm}{\protect\theoremname}
	  \else
      \newtheorem{thm}{\protect\theoremname}[chapter]
    \fi
\theoremstyle{plain}
    \ifx\thechapter\undefined
  \newtheorem{cor}{\protect\corollaryname}
\else
      \newtheorem{cor}{\protect\corollaryname}[chapter]
    \fi

\usepackage{hyperref}
\usepackage{authblk}

\newcommand{\blind}{0}

\newtheorem{Algorithm}{Algorithm}

\if0\blind
{
	\author{Aaron Fisher}
	\affil{Foundation Medicine Inc.; 150 Second St, Cambridge, MA 02141.
	\texttt{afishe27@alumni.jh.edu}.}
}
\fi

\if1\blind
{
	\author{[BLINDED]}
}
\fi

\DeclareMathOperator*{\argmin}{arg\,min}
\DeclareMathOperator*{\argmax}{arg\,max}

\usepackage{changepage}

\usepackage{xr}
\makeatletter
\newcommand*\bigcdot{\mathpalette\bigcdot@{.8}}
\newcommand*\bigcdot@[2]{\mathbin{\vcenter{\hbox{\scalebox{#2}{$\m@th#1\bullet$}}}}}
\makeatother
\newcommand{\bcd}{\bigcdot}

\usepackage{titlesec}

\makeatother

\providecommand{\corollaryname}{Corollary}
\providecommand{\theoremname}{Theorem}

\begin{document}
\title{Treatment Effect Bias from Sample Snooping:
Blinding Outcomes is Neither Necessary nor Sufficient}

\maketitle

\begin{abstract}
Popular guidance on observational data analysis states that outcomes
should be blinded when determining matching criteria or propensity
scores. Such a blinding is informally said to maintain the ``objectivity''
of the analysis, and to prevent analysts from fishing for positive
results by exploiting chance imbalances. Contrary to this notion,
we show that outcome blinding is not a sufficient safeguard against
fishing. Blinded and unblinded analysts can produce bias of the same
order of magnitude in cases where the outcomes can be \emph{approximately}
predicted from baseline covariates. We illustrate this vulnerability
with a combination of analytical results and simulations. Finally,
to show that outcome blinding is not necessary to prevent bias, we
outline an alternative sample partitioning procedure for estimating
the average treatment effect on the controls, or the average treatment
effect on the treated. This procedure uses all of the the outcome
data from all partitions in the final analysis step, but does not
require the analysis to not be fully prespecified. 
\end{abstract}

\section{Introduction\label{sec:Introduction}}

One of the central goals in statistics is to develop rigorous standards
of evidence in order to protect the conclusions of a study from being
influenced by the incentives or preconceptions of the researchers
involved. This goal arguably drives the popularity of prespecified
analysis plans \citep{mathieu2009comparison,humphreys2013fishing,gelman2013garden};
of p-value tests and objective Bayesian inference, in contrast to
subjective priors (see Chaper 13 of \citealp{Efron2016-bj}); and
of reproducible research practices concerning data management \citep{wickham2014tidy,wilkinson2016fair,weiskopf2017data}
and data analysis \citep{peng2021reproducible}. 

Within the field of causal inference, \emph{outcome blinding} forms
an additional, popular approach for generating credible evidence\emph{
}\citep{rubin2001using,rubin2007design}. This approach was\textcolor{black}{{}
partly motivated by the author's role as a litigation consultant for
the tobacco industry }\citep{rubin2001using,rubin2002ethics}\textcolor{black}{,
an industry whose objectivity is frequently called into question.}

\textcolor{black}{Under the outcome blinding workflow, researchers
are permitted to view covariate data while applying matching criteria,
propensity scores or other balancing techniques, but must finalize
their analysis plan }before observing outcome data. The rationale
offered for outcome blinding is that (1) viewing covariates allows
researchers to adjust their propensity score model if the covariate
distributions are still not well balanced across exposure groups after
weighting based on an initial model; (2) observational data analysis
should mimic randomized controlled trials, which generally require
the final analysis plan to be specified before the outcomes are observed;
and (3) masking outcomes prevents researchers from (subconsciously)
tinkering until a significant result is produced \citep{rubin2001using,rubin2007design,rubin2008objective}.

Outcome blinding has been widely endorsed and adopted over the past
two decades \citep{shadish2010primer,yue2012regulatory,yue2014designing,kainz2017improving,ding2017principal,lu2019good,king2019propensity_not_used,Chen2021-wl}.
In particular, outcome blinding was recently cited as a regulatory
necessity by the Real World Evidence Scientific Working Group, under
the Biopharmaceutical Section of the American Statistical Association
\citep{ho2021current}. This group's members include leading scientists
from across academia, industry and government.

Still, certain questions and debates about outcome blinding remain
open. \citeauthor{rubin2001using} \citeyearpar{rubin2001using,rubin2007design,rubin2008objective}
summarizes the benefits of outcome blinding by saying that it maintains
``objectivity,'' but this conceptualization of objectivity has not
yet been formalized as a property that can be proven or disproven.
The notion of mimicking randomized trials presents similar ambiguities,
as many randomized trials (e.g., those with rolling enrollment) require
committing to an analysis plan before seeing outcomes \emph{or covariates}.
Related questions also arise in discussions of propensity scores and
other balancing methods, which are commonly recommended because of
their compatibility with outcome blinding. Here, the effectiveness
of outcome blinding is implicitly assumed, but not directly investigated
\citep{steiner2010importance,li2016note,Levenson2021-ae}. Moreover,
blinding outcomes is not universally encouraged, as an inability to
view outcomes can impede confounder selection, limiting the effectiveness
of matching or weighting \citep{mccandless2009bayesian,de_luna2011covariate_selection,zigler2014uncertainty,shortreed2017outcome,damour2019_deconfounding_score}.
To our knowledge, deeper study of these issues is still desired \citep{varadhan2012review}.
This work aims to provide such a study. 

Namely, we demonstrate that outcome blinding does not preclude analysts
from fishing for desired results. Even a blinded analyst can  produce
intentionally biased estimators so long as some prior knowledge is
available that links covariates to outcomes (see discussion in Section
\ref{subsec:Problem-statement:-comparing}). We begin with a notation
framework for comparing the biases of blinded and unblinded analysts.
To form an initial, tractable model of analysts' behavior, we start
by assuming that analysts are fully aware of the underlying relationships
between covariates and outcomes in the population. We then show that,
in many scenarios, the resulting bias from blinded and unblinded analysts
is on the same order of magnitude. In sufficiently large samples,
the choices of these two analysts can even converge to each other.
Building on this base case, we show that similar results hold for
analysts who only have an approximate knowledge of the relationships
between covariates and outcomes (Section \ref{sec:Analytical-comparisons-of}).
We supplement these theoretical results with simulations over a wide
range of scenarios, which point to the same conclusion (Section \ref{sec:Simulated-comparisons-of}).

To highlight that outcome blinding is not necessary for preventing
bias, we discuss a sample partitioning method for estimating the average
treatment effect on the controls (ATC) or the average treatment effect
on the treated (ATT). This partitioning procedure avoids  bias without
``throwing away'' any of the outcome data from any of the partitions,
and without fully prespecifying all stages of the analysis (Section
\ref{sec:Sample-partitioning}). We close with a discussion (Section
\ref{sec:Discussion:-so-what?}). All proofs are provided in the supplementary
materials.

\section{Notation \& assumptions\label{subsec:Notation}}

We consider the scenario where analysts must choose one of several
treatment effect estimators. We will generally use boldface to denote
matrices or vectors, uppercase letters to denote random variables,
lowercase letters to denote realized values of random variables, and
a subscript $i$ to denote the index of the $i^{th}$ individual in
the sample.

Let $n$ denote the sample size, where each individual in the sample
is drawn independently from a common distribution. Let $\mathbf{X}$
denote a random $(n\times p$) matrix of covariates, where $\mathbf{X}_{i\bcd}$,
$\mathbf{X}_{\bcd j}$, and $\mathbf{X}_{ij}$ respectively denote
the $i^{th}$ row, the $j^{th}$ column, and the $i^{th}$ element
of the $j^{th}$ column. Let $\mathbf{\tilde{Y}}^{\text{treat}}$
and $\tilde{\mathbf{Y}}^{\text{control}}$ be $n$-length, latent,
random vectors representing potential outcomes on treatment and control
respectively. Here, higher values of $\tilde{\mathbf{Y}}^{\text{treat}}$
and $\tilde{\mathbf{Y}}^{\text{control}}$ denote better outcomes.
Let $\Delta=E(\tilde{Y}_{i}^{\text{treat}}-\tilde{Y}_{i}^{\text{control}})$
be the causal effect being estimated; let $\mathbf{A}\in\{0,1\}^{n}$
be a vector of treatment indicators; let $\mathbf{Y}=(Y_{1},\dots,Y_{n})$
be the observed outcomes, where $Y_{i}=A_{i}\tilde{Y}_{i}^{\text{treat}}+(1-A_{i})\tilde{Y}_{i}^{\text{control}}$;
and let $\mathcal{Y}$ and $\mathcal{X}$ denote the domain of $Y_{i}$
and $\mathbf{X}_{i\bcd}$ respectively. We assume throughout that
$0<P(A_{i}=1)<1$. 

Let $\mathcal{D}$ be a set of estimating functions for $\Delta$.
That is, let each $d\in\mathcal{D}$ be a mapping from $\text{\ensuremath{\mathcal{Y}}}^{n}\times\mathcal{X}^{n}\times\{0,1\}^{n}$
to the real line, such that $d(\mathbf{Y},\mathbf{X},\mathbf{A})$
produces an estimate of $\Delta$. For example, we may set $\mathcal{D}=\{d_{\text{OLS},1},\dots,d_{\text{OLS},p}\}$,
where $d_{\text{OLS},j}(\mathbf{y},\mathbf{x},\mathbf{a})$ returns
the ordinary least squares (OLS) coefficient from the linear regression
model that adjusts for the $j^{th}$ covariate $\mathbf{x}_{\bcd j}$:
\begin{align}
d_{\text{OLS},j}(\mathbf{y},\mathbf{x},\mathbf{a}) & =\argmin_{b_{a}}\left[\min_{b_{0},b_{j}}\vert\vert\mathbf{y}-\left(b_{0}+b_{j}\mathbf{x}_{\bcd j}+b_{a}\mathbf{a}\right)\vert\vert_{2}^{2}\right].\label{eq:d-def}
\end{align}
Here, $\{\mathbf{y},\mathbf{x},\mathbf{a}\}$ denotes a realization
of the random variables $\{\mathbf{Y},\mathbf{X},\mathbf{A}\}$. In
the results below, we will generally assume that each $d\in\mathcal{D}$
is linear in $\mathbf{Y}$. 

\subsection{Problem statement: comparing blinded and unblinded malicious analysts\label{subsec:Problem-statement:-comparing}}

This section introduces a notation framework to study the objectivity
arguments in Section \ref{sec:Introduction}. Analogous to the problem
framing used by \citet{king2019propensity_not_used}, we say that
an analyst is \emph{malicious} if they adaptively choose an estimator
$d$ in a way that maximizes bias. This means that an unblinded, malicious
analyst produces an estimate equal to 
\begin{equation}
\hat{\Delta}_{\text{snoop}}=\max_{d\in\mathcal{D}}d(\mathbf{Y},\mathbf{X},\mathbf{A}),\label{eq:snoop-def}
\end{equation}
after ``snooping'' by looking at the outcome data. Even if each
estimator $d\in\mathcal{D}$ is unbiased for $\Delta$, the maximum
across $\mathcal{D}$ will generally not be unbiased (see similar
discussion in \citealp{king2019propensity_not_used}). The larger
the set of candidates $\mathcal{D}$ is, the more opportunities the
analyst affords themself to exploit chance imbalances in pursuit of
a high estimated effect. Thus, by changing the size of the set $\mathcal{D}$,
we can represent analysts with different degrees of maliciousness. 

Unfortunately, similar bias can still exist\emph{ }even if the outcomes
are blinded,\emph{ }so long as they can be predicted. Suppose that
some of the variables in $\mathbf{X}$ are known to be correlated
with $\mathbf{Y}$, and let $\mu(\mathbf{X},\mathbf{A})=E(\mathbf{Y}|\mathbf{X},\mathbf{A})$
be the expected outcomes given the covariates and exposures. By replacing
$\mathbf{Y}$ with its proxy $\mu(\mathbf{X},\mathbf{A})$, a blinded
malicious analyst can select the estimator $D^{\star}\in\mathcal{D}$
that is \emph{expected} to produce the highest bias after the outcomes
are revealed. That is, they select the estimator function
\begin{align}
D^{\star} & =\argmax_{d\in\mathcal{D}}E_{\mathbf{Y}|\mathbf{X},\mathbf{A}}\left\{ d(\mathbf{Y},\mathbf{X},\mathbf{A})\right\} \nonumber \\
 & =\argmax_{d\in\mathcal{D}}d\left\{ \mu(\mathbf{X},\mathbf{A}),\mathbf{X},\mathbf{A}\right\} .\label{eq:worst-feature-blinded-x}
\end{align}
The analyst then commits to using $D^{\star}$ in their final analysis
of the outcomes, ultimately producing the estimate
\[
\hat{\Delta}_{\text{blind}}=D^{\star}(\mathbf{Y},\mathbf{X},\mathbf{A}).
\]
As with unblinded analysts, the extent of a blinded analyst's maliciousness
can be moderated by changing the size of the candidate set $\mathcal{D}$.
Above, Line (\ref{eq:worst-feature-blinded-x}) comes the linearity
of $d(\mathbf{y},\mathbf{x},\mathbf{a})$ with respect to $\mathbf{y}$.
The uppercase notation $D^{\star}$ captures the fact that the estimator
selection is random, as it depends on the random variables $\mathbf{A}$
and $\text{\emph{\ensuremath{\mathbf{X}}}}$. Hereafter, because the
terms $d\left\{ \mathbf{Y},\mathbf{X},\mathbf{A}\right\} $ and $d\left\{ \mu(\mathbf{X},\mathbf{A}),\mathbf{X},\mathbf{A}\right\} $
in Eqs (\ref{eq:snoop-def}) \& (\ref{eq:worst-feature-blinded-x})
differ only in their first argument, we will omit the dependence of
$d$ on $\mathbf{X},\mathbf{A}$. Instead, for any random variable
$\mathbf{Z}\in\mathcal{Y}^{n}$, we will write $d(\mathbf{Z})$ to
represent $d(\mathbf{Z},\mathbf{X},\mathbf{A})$, and will write $D^{\star}(\mathbf{Z})$
to represent $D^{\star}(\mathbf{Z},\mathbf{X},\mathbf{A})$. 

The conditional expectation $\mu(\mathbf{X},\mathbf{A})$ does not
always need to be particularly predictive. In some cases, we will
see that even weak associations between $\mathbf{X}$ and $\mathbf{Y}$
can have a dramatic effect on blinded bias (Section \ref{subsec:Overfitting-to-signal}). 

In many research settings, the function $\mu$ can be approximately
learned from previous studies, or from the related literature. This
is true, for example, in \citeauthor{rubin2001using}'s study of medical
expenses for smokers and similar nonsmokers \citep{rubin2001using,rubin2007design}.
Indeed, many scientific best practices implore quantitative researchers
to consult with domain experts before beginning an analysis \citep{vanderweele2019principles}.
If no existing domain expertise is available, initial pilot studies
are typically run in order to determine which variables are worth
measuring, and to estimate power for larger followup studies. As a
result of these pilot studies and consultations, analysts may already
know an approximation of $\mu$ before seeing the main dataset. 

Even in cases where no prior knowledge or pilot studies are available,
an analyst might still learn comparable information about $\mu$ via
sample splitting. That is, an analyst may reasonably choose to split
the data into two partitions, using one for informal exploration and
one for formal analysis. In the first partition, the analyst can attempt
to identify important prognostic variables (i.e., $\mu(\mathbf{X},\mathbf{A})$).
The second partition can then be used to estimate the effect of treatment.
However, even after the initial exploration, the analyst may opt to
continue to explore the distribution of covariates in the second data
partition before choosing an estimator, under the pretense of ensuring
that balance has been achieved. Thus, while balance checks would be
considered prudent under the outcome blinding paradigm (see Section
\ref{sec:Introduction}), they would also allow an ill-intentioned
analyst to overfit within the second partition, and to increase bias.

We close this section by noting that, as one reviewer has pointed
out, readers should be careful not to overinterpret the term ``outcome
blinding.'' This term is of course based on terminology from randomized
controlled trials (RCTs) in which patients and clinicians are blinded
to a patient's \emph{treatment.} However, while treatment randomization
hinders a RCT participant's ability to guess their treatment at baseline,
no comparable randomization can be done to prevent an analyst from
approximating participants' outcomes. For this reason, it could perhaps
be appropriate to rename ``outcome blinding'' as ``\emph{approximate}
outcome blinding,'' or even ``covariate-exposure previewing.''
Still, we keep the term ``outcome blinding'' for consistency with
\citeauthor{rubin2001using} \citeyearpar{rubin2001using,rubin2007design,rubin2008objective}.
The terminology of ``blinding'' also does not imply that no outcomes
have ever been observed from prior studies, as evidenced from \citeauthor{rubin2001using}'s
example of medical expenses \citep{rubin2001using,rubin2007design},
as well as the examples studied by \citet{rubin2008objective}.

\section{Analytical comparisons of blinded and snooping bias\label{sec:Analytical-comparisons-of}}

Next, we present analytical comparisons of the biases produced by
blinded an unblinded analysts. Our ultimate goal will be to offer
insight into the simulation results in Section \ref{sec:Simulated-comparisons-of},
below. 

Section \ref{subsec:Overfitting-to-signal}, examines large sample
settings, where, remarkably, the actions of blinded and unblinded
analysts can converge to each other. Section \ref{subsec:Finite-sample-bias},
examines the finite sample setting, where $E\left[\hat{\Delta}_{\text{blind}}\right]$
can be proportional to $E\left[\hat{\Delta}_{\text{snoop}}\right]$.
Section \ref{subsec:Analysts-with-imperfect} relaxes the assumption
that analysts know the conditional expectation function $E(\mathbf{Y}|\mathbf{X},\mathbf{A})$
exactly, and shows that versions of the previously discussed results
still hold under this relaxed setting. \textcolor{purple}{}

\subsection{Preference agreement in large samples \label{subsec:Overfitting-to-signal}}

In this section, beyond looking at each analyst's most preferred estimator,
we also consider their preference ordering for all of the available
estimators. For any two estimators $d,d'\in\mathcal{D}$, we say that
a snooping analyst \emph{prefers} $d$ to $d'$ if $d(\mathbf{Y})>d'(\mathbf{Y})$.
Likewise, we say that a blind analyst prefers $d$ to $d'$ if $d(\mu(\mathbf{X},\mathbf{A}))>d'(\mu(\mathbf{X},\mathbf{A}))$,
or, equivalently, if $E\left[d(\mathbf{Y},\mathbf{X},\mathbf{A})|\mathbf{X},\mathbf{A}\right]>E\left[d'(\mathbf{Y},\mathbf{X},\mathbf{A})|\mathbf{X},\mathbf{A}\right]$.
Our first result gives conditions under which the preferences of blinded
and snooping analysts converge to each other in large samples.
\begin{thm}
(Asymptotic preference agreement) Let $d_{j},d_{k}\in\mathcal{D}$
be two candidate estimators, where $j,k\in\{1,\dots,p\}$. Suppose
that the following conditions hold.\label{thm:signal-fitting}
\begin{enumerate}
\item \label{enu:OLS)}(OLS estimators) The two estimator functions $d_{j},d_{k}$
are equal to $d_{\text{OLS},j}$ and $d_{\text{OLS},k}$ respectively
(see Eq (\ref{eq:d-def})).
\item \label{enu:(Diffuse-signal)-for}\sloppy(Informative covariates)
$\mathbf{X}_{ij}$ and $\mathbf{X}_{ik}$ satisfy $\left|Cor(X_{ij},\mu(\mathbf{X}_{i\bcd},A_{i}))\right|,\left|Cor(X_{ik},\mu(\mathbf{X}_{i\bcd},A_{i}))\right|\in(0,1]$,
and $Cor(\mu(\mathbf{X}_{i\bcd},A_{i}),Y_{i})\in(0,1]$. 
\item \label{enu:(No-obs-confounders)-signal}(No observed confounders)
$\mathbf{X}\perp\mathbf{A}$. 
\end{enumerate}
Under these conditions,
\begin{equation}
P(d_{j}(\mathbf{Y})<d_{k}(\mathbf{Y})\text{\ensuremath{\text{ and }}}d_{j}(\mu(\mathbf{X},\mathbf{A}))>d_{k}(\mu(\mathbf{X},\mathbf{A}))\rightarrow0\label{eq:ranks-agree}
\end{equation}
as $n\rightarrow\infty$.
\end{thm}
It follows from Theorem \ref{thm:signal-fitting} that, if each estimator
in $\mathcal{D}$ takes the form of Eq (\ref{eq:d-def}) and Conditions
\ref{enu:(Diffuse-signal)-for}-\ref{enu:(No-obs-confounders)-signal}
hold for all $j,k\in\{1,\dots p\}$, then all preferences expressed
by snooping and blinded analysts will likely be identical in sufficiently
large samples. Thus, we would expect the selections of snooping and
blinded analysts to converge in large samples. This holds even if
these analysts do not always select the max estimator (e.g., Eq (\ref{eq:snoop-def})),
so long as their selection depends only on their preference ranks
for estimators in $\mathcal{D}$.

The intuition behind Theorem \ref{thm:signal-fitting} is that the
preferences of the snooping analyst depend on the chance associations
between features and treatment, as well as associations between features
and outcomes. Specifically, these preferences depend on $\frac{1}{n}\mathbf{A}^{\top}\text{\emph{\ensuremath{\mathbf{X}}}}$
and $\frac{1}{n}\mathbf{Y}^{\top}\text{\emph{\ensuremath{\mathbf{X}}}}.$
The first quantity $(\frac{1}{n}\mathbf{A}^{\top}\text{\emph{\ensuremath{\mathbf{X}}}}$)
is also available to the blinded analyst, and the second quantity
$(\frac{1}{n}\mathbf{Y}^{\top}\text{\emph{\ensuremath{\mathbf{X}}}})$
converges to the same limit as $\frac{1}{n}\mu(\mathbf{X},\mathbf{A})^{\top}\text{\emph{\ensuremath{\mathbf{X}}}}$.
Thus, in large samples, blinded analysts see information that is comparable
to what snooping analysts see, and so they can approximate the preferences
of snooping analysts.

While it is troubling to see how the behavior of our two analysts
align, we have not yet incorporated the effect of a potentially increasing
number of covariates $p$. This is a meaningful gap, as the bias of
both analysts will tend to decrease in large samples with fixed $p$,
as each estimator $d\in\mathcal{D}$ becomes less and less variable.
Thus, it will be important to be able to describe bias both in finite
samples and in samples with arbitrarily large $p$. These settings
are considered in the next subsection.

\subsection{Bias comparisons in finite samples\label{subsec:Finite-sample-bias}}

Here, we study the relative bias of $\hat{\Delta}_{\text{blind}}$
and $\hat{\Delta}_{\text{snoop}}$ in finite samples. Our main result
states that, on expectation, the blinded estimate can be proportional
to the snooping estimate. This result holds for any combination of
$n$ and $p$, and for any set of estimators $\mathcal{D}$ that are
linear in $\mathbf{Y}$.
\begin{thm}
\label{thm:noise-mix}(Finite sample expectation ratio) Let $\epsilon_{Y,i}=Y_{i}-\mu(\mathbf{X}_{i\bcd},A_{i})$,
let $\epsilon_{Y}=(\epsilon_{Y,1},\dots\epsilon_{Y,n})$, and let
$\rho=\sqrt{\frac{Var(\mu(\mathbf{X}_{i\bcd},A_{i}))}{Var(Y_{i})}}$
be the proportion of variance in outcomes that is explained by $\mathbf{X}$
and $\mathbf{A}$.

Suppose that
\begin{enumerate}
\item \label{enu:linearity_shift_invar}(Linearity) Each estimator function
$d_{j}(\mathbf{Y})$ is linear in \textbf{$\mathbf{Y}$} (see example
in Section \ref{subsec:Notation}).
\item \label{enu:noise-bad}(Higher signal-to-noise ratios increase bias)
We assume that
\[
E\left[\text{\ensuremath{\max_{d\in\mathcal{D}}}}\,\,\,d\left(\frac{\mathbf{Y}}{\text{Var}\left(Y_{i}\right)^{1/2}}\right)\right]\leq E\left[\text{\ensuremath{\max_{d\in\mathcal{D}}\,\,\,d\left(\frac{\mathbf{Y}-\epsilon_{Y}}{\text{Var}\left(Y_{i}-\epsilon_{Y,i}\right)^{1/2}}\right)}}\right].
\]
That is, the more that variance in Y can be explained by X, the higher
the unblinded bias will tend to be (see discussion below).
\end{enumerate}
Under these conditions,
\begin{equation}
\frac{E\left[\hat{\Delta}_{\text{blind}}\right]}{E\left[\hat{\Delta}_{\text{snoop}}\right]}\geq\rho.\label{eq:bias-ratio}
\end{equation}
\end{thm}
Eq (\ref{eq:bias-ratio}) states that, on expectation, the blinded
estimate is at least proportional to the snooping estimate, with a
proportionality constant ($\rho$) that depends on how well $\mathbf{Y}$
can be predicted from $\mathbf{X}$. For example, any addition to
the candidate set $\mathcal{D}$ will necessarily increase the bias
of a snooping analyst, but it will also lead to proportionate increase
in the bias of a blinded analyst. In other words, in cases where
the bias caused by viewing outcomes is large, Theorem \ref{thm:noise-mix}
suggests that the bias caused by merely observing $\mathbf{X}$ and
$\mathbf{A}$ can be large as well.

Roughly speaking, Condition \ref{enu:noise-bad} means that the noisier
the relationship is between $\mathbf{X}$ and $\mathbf{Y}$, the harder
it is for an unblinded analyst to amplify the treatment effect by
exploiting chance imbalances. We expect this condition to hold in
many cases, as adjusting for a chance imbalance in any given covariate
$\mathbf{X}_{\bcd j}$ is relatively ineffectual when $\mathbf{X}_{\bcd j}$
is not prognostic of $\mathbf{Y}$. For example, consider the estimator
$d_{\text{OLS},j}$ that linearly adjusts for $\mathbf{X}_{\bcd j}$.
The value of $|d_{\text{OLS},j}(\mathbf{Y},\mathbf{X},\mathbf{A})|$
depends on four factors: the unadjusted average difference in $\mathbf{Y}$
across treatment arms; the imbalance in $\mathbf{X}_{\bcd j}$ across
treatment arms; the scale of $\mathbf{Y}$; and the correlation between
$Y_{i}$ and $X_{ij}$. For $|d_{\text{OLS},j}(\mathbf{Y},\mathbf{X},\mathbf{A})|$
to be large, at least one of these four factors must be large. Increasing
the signal-to-noise ratio in the manner of Condition \ref{enu:noise-bad}
(i.e., removing noise) will increase the correlation between $Y_{i}$
and $X_{ij}$, but leave the first three factors unchanged. Thus,
if $\mathcal{D}=\{d_{\text{OLS},1},\dots,d_{\text{OLS},p}\}$, then
we would expect higher signal-to-noise ratios to be associated with
larger values of $E\max_{d\in\mathcal{D}}d(\mathbf{Y})$. Indeed,
Condition \ref{enu:noise-bad} held over the wide range of simulation
settings we explored in Section \ref{sec:Simulated-comparisons-of},
below.

Alternatively, in the special case where Condition \ref{enu:noise-bad}
holds with equality, Eq (\ref{eq:bias-ratio}) also holds with equality
(see Section \ref{subsec:Proof-of-Theorem-noise} of the supplementary
materials). This case is especially relevant when $\mathbf{X}$ contains
a large number of ``noisy'' covariates that are independent of $\mu(\mathbf{X},\mathbf{A})$.
For example, suppose that $\mathcal{D}=\{d_{\text{OLS},1},\dots,d_{\text{OLS},p}\}$,
such that analysts select a single covariate to adjust for. Suppose
also that $\mu(\mu(\mathbf{X}_{i\bcd},A_{i})$ and $\epsilon_{Y,i}$
are both normal variables with mean zero. Here, removing signal from
$\mathbf{Y}$ will only affect the estimators $d_{\text{OLS},j}(\mathbf{Y})$
for which $\mathbf{X}_{\bcd j}$ is predictive of $\mathbf{Y}$. If
almost none of the covariates are predictive of $\mathbf{Y}$, then
removing noise from $\mathbf{Y}$ in the manner of Condition \ref{enu:noise-bad}
will have minimal effect on $\max_{d\in\mathcal{D}}d(\mathbf{Y})$,
since signal variables will rarely be selected anyway. As a result,
Condition \ref{enu:noise-bad} and Eq (\ref{eq:bias-ratio}) will
both hold with approximate equality. We will refer back to this special
case when discussing simulation results.

\subsection{Analysts with imperfect knowledge of $E(\mathbf{Y}|\mathbf{X},\mathbf{A})$\label{subsec:Analysts-with-imperfect}}

In this section we relax assumption that analysts know the conditional
expectation function $\mu(\mathbf{X},\mathbf{A})=E(\mathbf{Y}|\mathbf{X},\mathbf{A})$
exactly, and instead assume that they know only a rough approximation
denoted by $\hat{\mu}$. We will refer to blinded analysts who base
their decisions on the imperfect proxy function $\hat{\mu}$ as \emph{misinformed.}
The approximation $\hat{\mu}$ can be determined from an pilot study,
from consultations with domain experts, or from a sample splitting
procedure (see Section \ref{subsec:Problem-statement:-comparing}).
We will see below that Theorems \ref{thm:signal-fitting} \& \ref{thm:noise-mix}
can be extended to describe bias of misinformed analysts, and that
degree to which such analysts can produce bias depends on the fidelity
of the approximation $\hat{\mu}$.

Before discussing these results, we first introduce more precise notation
to describe the behavior of a misinformed analyst. Given a prespecified
approximation $\hat{\mu}$, let $D_{\text{\ensuremath{\hat{\mu}}}}^{\star}=\argmax_{d\in\mathcal{D}}d\left(\hat{\mu}(\mathbf{X},\mathbf{A})\right)$
be the estimator that the misinformed analyst believes will be the
largest after\textbf{ }$\mathbf{Y}$ is revealed, and let $\hat{\Delta}_{\text{blind},\hat{\mu}}=D_{\text{\ensuremath{\hat{\mu}}}}^{\star}\left(\mathbf{Y}\right)$
be their resulting estimate. Let $\epsilon_{(d)}=d\left(\hat{\mu}(\mathbf{X},\mathbf{A})\right)-d\left(\mu(\mathbf{X},\mathbf{A})\right)$
be the error that a misinformed analyst incurs by not knowing $\mu$,
for a given estimation function $d$. It is straightforward to show
that, if $d(\mathbf{Y})$ is linear in $\mathbf{Y}$ for each $d\in\mathcal{D}$,
then
\begin{equation}
E\left[\hat{\Delta}_{\text{blind},\hat{\mu}}\right]\geq E\left[\hat{\Delta}_{\text{blind}}\right]-E\left[\epsilon_{(D_{\hat{\mu}}^{\star})}-\epsilon_{(D^{\star})}\right],\label{eq:mu-bias-diff}
\end{equation}
where $D^{\star}$ is the function chosen by a blinded analyst with
perfect knowledge of $\mu$ (see Eq (\ref{eq:worst-feature-blinded-x})
and Section \ref{subsec:Proof-of-Eq-bias-mu-diff} of the supplementary
materials). That is, the difference between the bias of a misinformed
blinded analyst and the bias of a blinded analyst who knows the true
function $\mu$ depends on the accuracy of the proxy function $\hat{\mu}$
(via $E\left[\epsilon_{(D_{\hat{\mu}}^{\star})}-\epsilon_{(D^{\star})}\right]$).

Under assumptions similar to those used in Section \ref{subsec:Finite-sample-bias},
the next corollary simplifies Eq (\ref{eq:mu-bias-diff}) into a more
interpretable statement about the extent to which misinformation about
$\mu$ reduces an analyst's ability to bias their results.
\begin{cor}
\label{cor:mu-finite-bias-ratio}Let $\hat{\mu}(\mathbf{X},\mathbf{A})=\mu(\mathbf{X},\mathbf{A})+\mathbf{W}$
be noise-corrupted version of $\mu(\mathbf{X},\mathbf{A})$ that is
available to a misinformed, malicious, blinded analyst, where $\mathbf{\mathbf{W}}=(W_{1},\dots,W_{n})$
is a vector of random prediction errors. 

Suppose that the following conditions hold.
\begin{enumerate}
\item \label{enu:(mu-Linearity)-Each-}(Linearity) $d(\mathbf{Y})$ is linear
in $\mathbf{Y}$ for each $d\in\mathcal{D}$.
\item \label{enu:(mu-Independent-errors)-.}(Independent, homogeneous errors)
$\mathbf{W}\perp\mathbf{Y},\mathbf{X},\mathbf{A}$.
\item \label{enu:(mu-Common-distributions-up-1}(Normality) $\mathbf{W}$
and $\mu(\mathbf{X},\mathbf{A})$ are both normally distributed with
mean zero.
\item \label{enu:(mu-Noise-does-not}(Higher signal-to-noise ratios increase
bias) $E\left[\max_{d\in\mathcal{D}}d\left(\text{\textbf{U}}\right)\right]\leq E\left[\max_{d\in\mathcal{D}}d\left(\mu(\mathbf{X},\mathbf{A})\right)\right]$,
where $\mathbf{U}$ is an independent copy of $\mu(\mathbf{X},\mathbf{A})$
with the same marginal distribution, but with $\mathbf{U}\perp\mathbf{X},\mathbf{A}$.
\end{enumerate}
Under the above conditions
\begin{equation}
\frac{E\left[\hat{\Delta}_{\text{blind},\hat{\mu}}\right]}{E\left[\hat{\Delta}_{\text{blind}}\right]}\geq1-2\sqrt{\frac{\text{Var}(W_{i})}{\text{Var}(\mu(\mathbf{X}_{i\bcd},A_{i}))}}.\label{eq:mu-ratio}
\end{equation}
\end{cor}
Eq (\ref{eq:mu-ratio}) states that the extent to which bias is reduced
by misinformation (the left-hand side) is limited by the magnitude
of misinformation (as represented by $\text{Var}(W_{i})$). As $\text{Var}(W_{i})$
approaches zero, $\hat{\mu}$ becomes more accurate, and the expectation
of the estimator from a blinded, misinformed analyst approaches the
expectation of the estimator from a blinded analyst with full knowledge
of $E(\mathbf{Y}|\mathbf{X},\mathbf{A})$. The required Condition
\ref{enu:(mu-Noise-does-not} is analogous to Condition \ref{enu:noise-bad}
of Theorem \ref{thm:noise-mix} (see discussion in Section \ref{subsec:Finite-sample-bias}).

Similarly, Theorem \ref{thm:signal-fitting} can also be extended
to the setting where $\mu$ is unknown, but where an increasingly
accurate proxy $\hat{\mu}$ is available. Suppose that $\hat{\mu}$
is learned from an independent pilot study of size $n_{\text{train}}$,
such that $\hat{\mu}$ becomes more accurate as $n_{\text{train}}$
increases. Equivalently, given a sufficient amount of data, $\hat{\mu}$
can be learned using a sample splitting approach. For and two functions
$d,d'\in\mathcal{D}$, we will say that the blinded, misinformed analyst
prefers $d$ to $d'$ if $d(\hat{\mu}(\mathbf{X},\mathbf{A}))\geq d'(\hat{\mu}(\mathbf{X},\mathbf{A}))$.
A simple corollary of Theorem \ref{thm:signal-fitting} is that, if
$\hat{\mu}$ approaches $\mu$ in large samples, then the preference
convergence shown in Eq (\ref{eq:ranks-agree}) still holds.
\begin{cor}
\label{cor:signal-fitting-proxy} (Preference agreement with learned
outcome proxies) If the conditions of Theorem \ref{thm:signal-fitting}
hold and $E\left[\left\{ \hat{\mu}(\mathbf{X}_{i\bcd},A_{i})-\mu(\mathbf{X}_{i\bcd},A_{i})\right\} ^{2}\right]\rightarrow0$
as $n_{\text{train}}\rightarrow\infty$, then

\begin{equation}
P(d_{j}(\hat{\mu}(\mathbf{X},\mathbf{A}))<d_{k}(\hat{\mu}(\mathbf{X},\mathbf{A})\text{ and }d_{j}(\mathbf{Y})>d_{k}(\mathbf{Y}))\rightarrow0\label{eq:ranks-agree-1}
\end{equation}
as $n,n_{\text{train}}\rightarrow\infty$.
\end{cor}
In words, Eq (\ref{eq:ranks-agree-1}) states that, when given enough
data, a blinded analyst who can effectively learn $\mu(\mathbf{X},\mathbf{A})$
will likely make decisions that are similar to those of snooping analysts. 

\section{Simulated comparisons of blinded and unblinded bias\label{sec:Simulated-comparisons-of}}

While the analytical results in Section \ref{sec:Analytical-comparisons-of}
paint a worrisome picture of blinded bias, several questions warrant
deeper exploration. For Theorem \ref{thm:signal-fitting}, we have
not yet examined how quickly the blinded analyst's behavior converges
to that of the snooping analyst, especially when the blinded analyst
must learn the conditional expectation function $\mu$ (see Corollary
\ref{cor:signal-fitting-proxy}). For Theorem \ref{thm:noise-mix},
we have not examined how conservative the lower bound on $E\left[\hat{\Delta}_{\text{blind}}\right]$
is (Eq (\ref{eq:bias-ratio})). It will also be beneficial to check
how often the signal-to-noise condition in Theorem \ref{thm:noise-mix}
holds (Condition \ref{enu:noise-bad} of Theorem \ref{thm:noise-mix}). 

\sloppy To inform these questions, we simulate data under a variety
of settings. We set $\mathcal{D}=\{d_{\text{OLS},1},\dots,d_{\text{OLS},p}\}$
(see Eq (\ref{eq:d-def})) and set $Y=\mathbf{X}\beta+\epsilon$,
but vary the sample size $(n),$ the sample dimension ($p$), and
the variance in $Y_{i}$ explained by $\mathbf{X}_{i\bcd}$ ($\rho^{2}=Var(\mu(\mathbf{X}_{i\bcd},0))/Var(Y_{i}-A_{i}\Delta)$).
Here, each row $\mathbf{X}_{i\bcd}$ follows an independent, standard
normal distribution, regardless of the value of $A_{i}$; $\beta=(\beta_{1},\beta_{2},\dots,\beta_{p})$
is a $p$-length vector with $\beta_{j}=2$ for $j\in[1,5]$, $\beta_{j}=-1$
for $j\in[6,10]$, and $\beta_{j}=0$ for $j>10$; and $\epsilon_{i}$
is normally distributed with mean zero and variance calibrated to
achieve the desired value for $\rho^{2}$. To avoid degenerate estimates
in small samples, we fix\textbf{ $\mathbf{A}$ }so that the two treatment
arms are the same size (i.e., we set $A_{i}=1(i<n/2)$). Under this
model, we simulate all combinations of $n\in\{30,100,250,500\}$;
$p\in\{10,30,100,500\}$; $\rho^{2}\in\{0.25,0.5,0.75\}$. In each
setting, we simulate 2500 draws. 

We also consider two levels of knowledge regarding the conditional
expectation function $\mu$. In the first setting, we assume $\mu$
is known to the blinded analyst. In the second setting, we assume
that the blinded analyst estimates $\mu$ by fitting a lasso regression
on an independent dataset of size $n_{\text{train}}=n$ (see Section
\ref{subsec:Analysts-with-imperfect}), with a penalty parameter determined
from cross-validation \citep{Friedman2010lasso}. 

\begin{figure}
\begin{raggedleft}
\includegraphics[width=0.97\columnwidth]{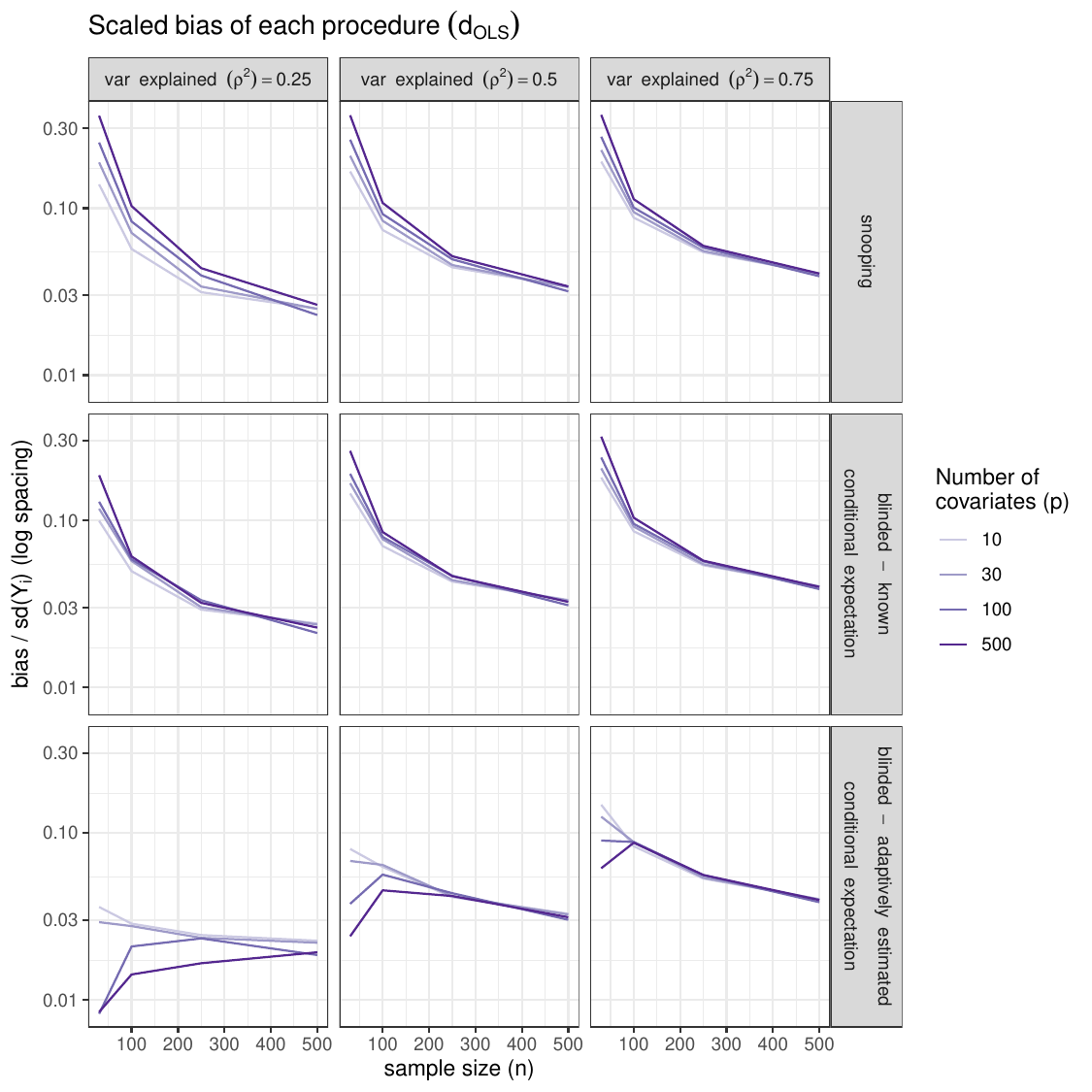}
\par\end{raggedleft}
\caption{\label{fig:simulations-OLS-scaled}Simulated bias from snooping and
blinded analysts using $d_{\text{OLS}}$ -- Columns of plots show
different values for the variance in outcomes that is explained by
covariates, $\rho^{2}=Var(\mu(\mathbf{X}_{i\bcd},0))/Var(Y_{i}-A_{i}\Delta)$.
The first row of plots shows the bias of snooping analysts. The second
row shows the bias of blinded analysts who know the conditional expectation
function $\mu(\mathbf{X},\mathbf{A})$ a priori. The third row shows
the bias of blinded analysts who must estimate $\mu(\mathbf{X},\mathbf{A})$
from an independent training sample (see Section \ref{subsec:Analysts-with-imperfect}).
The y-axis shows the bias of each analyst, scaled by the standard
deviation of $Y_{i}$. This scaling is done to account for the fact
that $Var(\epsilon_{i})$ changes in order to achieve each desired
value for $\rho^{2}$.}
\end{figure}

\begin{figure}
\begin{raggedleft}
\includegraphics[width=0.97\columnwidth]{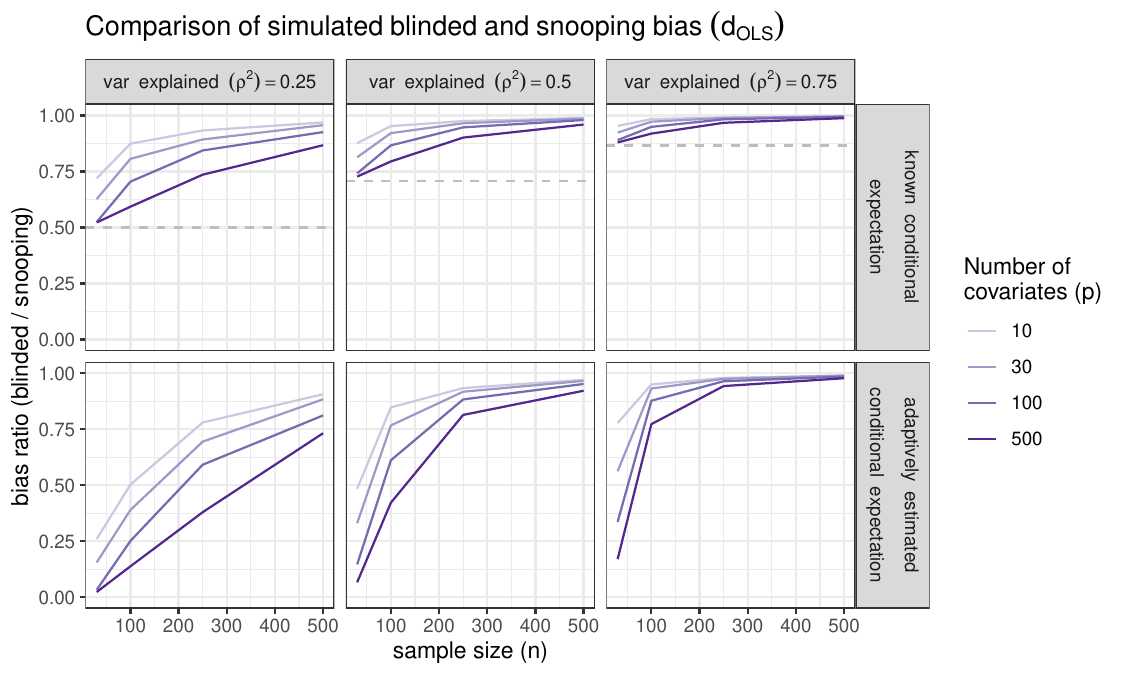}
\par\end{raggedleft}
\caption{\label{fig:simulations-OLS-ratio}Comparison of bias from simulated
snooping and blinded analysts, each using $d_{\text{OLS}}$ -- Columns
of plots show different values for the variance in outcomes that is
explained by covariates, $\rho^{2}=Var(\mu(\mathbf{X}_{i\bcd},0))/Var(Y_{i}-A_{i}\Delta)$.
Rows of plots indicate whether or not $\mu(\mathbf{X},\mathbf{A})$
is known to the blinded analyst, or must be estimated from an independent
training sample (see Section \ref{subsec:Analysts-with-imperfect}).
The y-axis shows the bias ratio between blinded and snooping analysts,
$E\left[\hat{\Delta}_{\text{blind}}-\Delta\right]/E\left[\hat{\Delta}_{\text{snoop}}-\Delta\right]$.
Since $\Delta=0$ in our simulations, this bias ratio is also equal
to the ratio of expectations described in Theorem \ref{thm:noise-mix}.
For cases where $\mu(\mathbf{X},\mathbf{A})$ is known (top row),
the dashed, gray line shows the lower bound on the bias ratio suggested
by Theorem \ref{thm:noise-mix} ($\rho$).}
\end{figure}

Figures \ref{fig:simulations-OLS-scaled} \& \ref{fig:simulations-OLS-ratio}
display the results of our simulations. Figure \ref{fig:simulations-OLS-scaled}
shows the bias for three types of analysts: (1) snooping analysts
who observe $\mathbf{Y}$, (2) blinded analysts who know the function
$\mu(\mathbf{X},\mathbf{A})$ a priori, and (3) blinded analysts who
estimate $\mu(\mathbf{X},\mathbf{A})$ using an independent dataset.
In the first two settings, bias decreases as $n/p$ increases. This
is as expected, since each estimate $d_{\text{OLS},j}(\mathbf{Y})$
approaches zero in large samples. The trend is more complex in the
third setting however, where larger values of $n/p$ also facilitate
learning the function $\mu$. Across all settings we considered, changes
in $n$ generally produced more dramatic effects than changes in $p$.

\sloppy To explicitly compare blinded and snooping analysts, Figure
\ref{fig:simulations-OLS-ratio} displays the bias ratio $E\left[\hat{\Delta}_{\text{blind}}-\Delta\right]/E\left[\hat{\Delta}_{\text{snoop}}-\Delta\right]$.
That is, Figure \ref{fig:simulations-OLS-ratio} divides the second
and third rows of Figure \ref{fig:simulations-OLS-scaled} by the
first row of Figure \ref{fig:simulations-OLS-scaled}. Since $\Delta=0$
in our simulations, the bias ratio depicted in Figure \ref{fig:simulations-OLS-ratio}
is also equal to the ratio of expectations described in Eq (\ref{eq:bias-ratio}).
This ratio is substantial in many cases. For example, in settings
where $\mu$ must be estimated and $n=n_{\text{train}}\geq100$, we
observe that the bias ratio is similar to or larger than $\rho^{2}$. 

When $\mu$ is known a priori, the bias ratio in Figure \ref{fig:simulations-OLS-ratio}
is at least $\rho$ (dashed horizontal line), as suggested by Theorem
\ref{thm:noise-mix}. Condition \ref{enu:noise-bad} of Theorem \ref{thm:noise-mix}
is difficult to confirm analytically, but was satisfied empirically
in every simulation setting we considered. The lower bound in Theorem
\ref{thm:noise-mix} appears to hold with approximate equality only
when $n<<p$, and to be conservative otherwise. This trend is consistent
with our result in Section \ref{subsec:Overfitting-to-signal} stating
that blinded and unblinded analysts tend to agree in large samples,
as well as with the discussion at the end of Section \ref{subsec:Finite-sample-bias}.

In addition to the simulations described above, we also implemented
estimators based on inverse propensity score weighting (IPW). Like
the OLS estimators described in this section, each IPW estimator adjusted
for a different covariate. The results were almost identical, and
are described in detail in the supplementary materials. Condition
\ref{enu:noise-bad} of Theorem \ref{thm:noise-mix} held in each
of these simulation settings as well.

\section{Outcome blinding is not necessary: other methods to avoid bias\label{sec:Sample-partitioning}}

Thus far, we have shown that outcome blinding is not sufficient for
preventing bias. In this section, we discuss simple strategies that
\emph{can} prevent the bias incurred by adaptively selecting an estimator.
Of course, one valid, conventional approach is to determine an estimator
from a subsample and apply this estimator in a separate subsample
to estimate treatment effects. Unfortunately, this requires that we
``throw away'' half of the outcome data in the final estimation
step.

As a related approach, we now introduce a procedure that applies unblinded
sample splitting while still allowing all data points to be used in
the final treatment effect estimation, so long as the final estimand
is either the ATT or the ATC. As an illustration, we focus on estimating
the ATT:
\[
E(\tilde{Y}_{i}^{\text{treat}}-\tilde{Y}_{i}^{\text{control}}|A_{i}=1)=E(\tilde{Y}_{i}^{\text{treat}}|A_{i}=1)-E(\tilde{Y}_{i}^{\text{control}}|A_{i}=1).
\]
We will see that, while our proposed procedure does not reproduce
all properties of traditional sample splitting, it is sufficient to
allow data exploration without incurring bias.

In the same way that $\mathcal{D}$ denotes a set of estimators for
$\Delta$, let $\mathcal{F}$ denote a set of estimators for $E(\tilde{Y}_{i}^{\text{control}}|A_{i}=1)$.
That is, if $f\in\mathcal{F}$ and $\mathcal{I}\subseteq\{1,\dots n\}$,
then $f\left(\{Y_{i},\mathbf{X}_{i\bcd},A_{i}\}_{i\in\mathcal{I}}\right)$
represents an estimate of $E(\tilde{Y}_{i}^{\text{control}}|A_{i}=1)$.
This set $\mathcal{F}$ may be arbitrarily large, or even infinite
in size. Different estimators in $\mathcal{F}$ may be used to represent
screening out different subsets of covariates, different methods for
feature generation, or different styles of parametric modeling (see,
for example, \citealp{lee2010improving,hill2011bayesian,wang2017g,dorie2019automated}). 

Given a set of candidate estimator functions $\mathcal{F}$, we propose
the following interactive procedure.

\begin{Algorithm}(Sample splitting for the ATT)\label{alg:Sample-splitting-for}
\begin{enumerate}
\item (Partition) The researcher partitions the treated data into two parts,
$\mathcal{I}_{1}^{\text{treat}}$ and $\mathcal{I}_{2}^{\text{treat}}$.
\label{enu:(Partition)}
\item \label{enu:(Explore)-Freely-explore}(Explore) The researcher freely
explores $\{Y_{i}^{\text{}},\mathbf{X}_{i\bcd}^{\text{}}\}_{i\in\mathcal{I}_{1}^{\text{treat}}}$.
Based on insights from this exploration, as well as on prior knowledge,
the researcher uses their personal judgement to select a function
$f\in\mathcal{F}$ for estimating $E(\tilde{Y}_{i}^{\text{control}}|A_{i}=1)$.
Up until this point, the researcher has only had access to data from
$\mathcal{I}_{1}^{\text{treat}}$.
\item \label{enu:(Fit)-Apply-this}(Fit) After committing to a particular
estimator $f\in\mathcal{F}$, the researcher applies $f$ to the remaining
data from $\mathcal{I}_{2}^{\text{treat}}$ and from the control arm,
and returns the following estimate of the ATT.
\begin{align}
\hat{\Delta}_{\text{ATT},f}=\left(\frac{\sum_{i=1}^{n}Y_{i}A_{i}}{\sum_{i=1}^{n}A_{i}}\right)-f\left(\{Y_{i},\mathbf{X}_{i\bcd},A_{i}\}_{\{i\,:\,i\in\mathcal{I}_{2}^{\text{treat}}\text{ or }A_{i}=0\}}\right).\label{eq:fit-est}
\end{align}
\end{enumerate}
\end{Algorithm}Before discussing formal properties of this algorithm,
we illustrate the steps above using a hypothetical data example. 

\subsection{Example application\label{subsec:Example-application}}

Consider a hypothetical, observational study of vaccine efficacy.
For a given individual $i$, let $A_{i}$ represent their exposure
to a vaccine, let $X_{i}$ denote their age, and let $Y_{i}$ be an
indicator of infection within 1 year. Let $n=1000$ be the number
of study participants. 

\begin{figure}

\begin{centering}
\includegraphics[width=0.7\columnwidth]{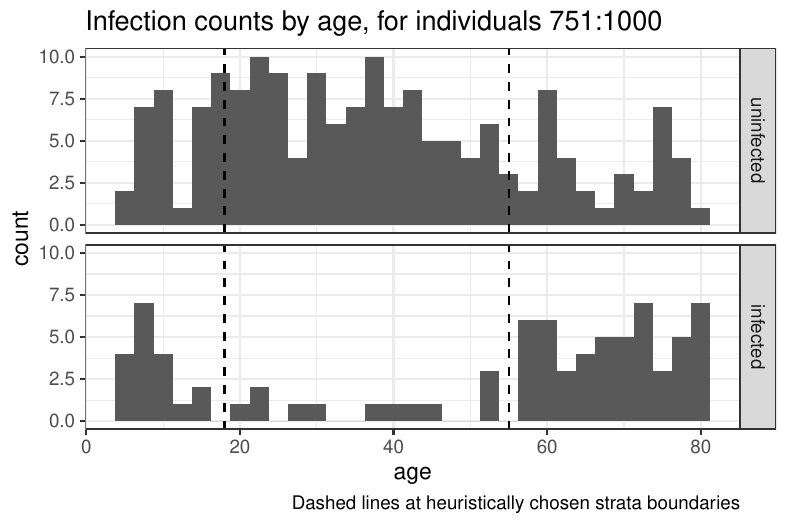}\caption{\label{fig:Example-exploratory-analysis}Example exploratory analysis
-- Infection status by age for participants $i\in\{751,752,\dots,1000\}=\mathcal{I}_{\text{treat}}^{1}$
in the hypothetical study outlined in Section \ref{subsec:Example-application}.}
\par\end{centering}
\end{figure}

The following workflow exemplifies how one might implement Algorithm
\ref{alg:Sample-splitting-for} to estimate the average effect of
the vaccine on those who received it. First, the researcher defines
a set of candidate estimators $\mathcal{F}$. In this case, they choose
the set of stratified estimators formed by using no more than 10 age-specific
strata (example to follow). Next, the researcher defines data partitions
(Step \ref{enu:(Partition)}). For ease of notation, we will assume
that half of the study participants received the vaccine, with $A_{i}=0$
for $i\leq500$ and $A_{i}=1$ for $i>500$. Suppose that the researcher
sets $\mathcal{I}_{2}^{\text{treat}}=\{501,502,\dots750\}$, and $\mathcal{I}_{1}^{\text{treat}}=\{751,752,\dots,1000\}$.
The researcher then performs an exploratory analysis of data from
participants in $\mathcal{I}_{\text{treat}}^{1}$ by plotting infections
status against age, producing Figure \ref{fig:Example-exploratory-analysis}
(Step \ref{enu:(Explore)-Freely-explore}). Upon seeing this plot,
the researcher informally concludes that younger children and older
adults have higher infection risks, and so they decide to stratify
individuals according to three age intervals: $(0,18],(18,55]$, and
$(55,\infty)$. That is, the researcher commits to use strata boundaries
equal to $\{u_{1},u_{2},u_{3},u_{4}\}=\{0,18,55,\infty\}$, and to
estimate $E(\tilde{Y}_{i}^{\text{control}}|A_{i}=1)$ using the stratified
estimator
\begin{align}
 & f_{18,55}\left(\{Y_{i},X_{i},A_{i}\}_{i=1}^{750}\right)\nonumber \\
 & \hspace{1em}=\sum_{k=2}^{4}\left\{ \frac{\sum_{i=1}^{750}Y_{i}(1-A_{i})1(X_{i}\in(u_{k-1},u_{k}])}{\sum_{i=1}^{750}(1-A_{i})1(X_{i}\in(u_{k-1},u_{k}])}\times\frac{\sum_{i=1}^{750}A_{i}1(X_{i}\in(u_{k-1},u_{k}])}{\sum_{i=1}^{750}A_{i}}\right\} .\label{eq:strat}
\end{align}
Note that, above, $\{1,\dots,750\}$, is equivalent to the set $\{i\,:\,i\in\mathcal{I}_{2}^{\text{treat}}\text{ or }A_{i}=0\}$
mentioned in Algorithm \ref{alg:Sample-splitting-for}. Each summation
term in Eq \ref{eq:strat} involves taking the mean control outcome
in one of the age strata, and weighting it according to the prevalence
of that strata among the treated. 

Finally, the researcher returns
\begin{equation}
\left(\frac{\sum_{i=1}^{1000}Y_{i}A_{i}}{\sum_{i=1}^{1000}A_{i}}\right)-f_{18,55}\left(\{Y_{i},X_{i},A_{i}\}_{i=1}^{750}\right)\label{eq:ex-est}
\end{equation}
as their estimate of the ATT (Step \ref{enu:(Fit)-Apply-this}). Importantly,
the first term in Eq \ref{eq:ex-est} was unaffected by the qualitative
choices made by the researcher in selecting $f_{18,55}$, and the
researcher's selection was required to be completed before they had
access to the data on which $f_{18,55}$ was applied ($\{Y_{i},X_{i},A_{i}\}_{i=1}^{750}$,
in Step \ref{enu:(Fit)-Apply-this}).

\subsection{\label{subsec:Properties-of-Algorithm}Properties of Algorithm \ref{alg:Sample-splitting-for}}

All outcomes from the dataset are used in the final estimate from
Algorithm \ref{alg:Sample-splitting-for}, in Eqs (\ref{eq:fit-est})
\& (\ref{eq:ex-est}). However, the first term in Eq (\ref{eq:fit-est})
is not influenced by the exploratory procedures in Step \ref{enu:(Explore)-Freely-explore}.
Likewise, the data used to select $f$ is independent of the data
that is plugged into $f$, in the second term. As we will see below,
this means that the exploratory steps do not inject additional bias.
Roughly speaking, an analyst who learns about the data generating
distribution from exploratory analyses, using Algorithm \ref{alg:Sample-splitting-for},
will be no more biased than an analyst who learns the same information
from external sources.

To formalize this statement, we will treat the researcher's choice
of estimator as a random variable, denoted by $F_{\text{split}}\in\mathcal{F}$
(Step \ref{enu:(Explore)-Freely-explore}). That is, $F_{\text{split}}$
is a random selection from $\mathcal{F}$ that can depend only on
$\{Y_{i}^{\text{}},\mathbf{X}_{i\bcd}^{\text{}}\}_{i\in\mathcal{I}_{1}^{\text{treat}}}$.
In our notation, we will also take special care to account for the
fact that the partition itself, $\mathcal{I}_{1}^{\text{treat}}\subseteq\{1,\dots,n\}$,
is also a random variable, as it depends on $\mathbf{A}$. 

By definition, we know that $F_{\text{split}}\perp\{Y_{i},\mathbf{X}_{i\bcd},A_{i}\}_{i\not\in\mathcal{I}_{1}^{\text{treat}}}|\mathcal{I}_{1}^{\text{treat}}$.
From this conditional independence statement, it follows immediately
that 
\begin{equation}
E\left(\hat{\Delta}_{\text{ATT},F_{\text{split}}}|\mathcal{I}_{1}^{\text{treat}},F_{\text{split}}=f\right)=E\left(\hat{\Delta}_{\text{ATT},f}|\mathcal{I}_{1}^{\text{treat}}\right)\label{eq:att-same}
\end{equation}
for any $f\in\mathcal{F}$. Roughly speaking, this means if a researcher
following Algorithm \ref{alg:Sample-splitting-for} selects $F_{\text{split}}=f$,
then the expectation of their resulting estimate $\hat{\Delta}_{\text{ATT},f}$
can be interpreted \emph{as if} the function $f$ were chosen a priori,
based on only the indices $\mathcal{I}_{1}^{\text{treat}}\subseteq\{1,\dots,n\}$
but not the values $\{Y_{i},\mathbf{X}_{i\bcd}\}_{i\in\mathcal{I}_{1}^{\text{treat}}}$
associated with those indices. For example, in the scenario described
in Section \ref{subsec:Example-application}, knowing the indices
$\mathcal{I}_{1}^{\text{treat}}\subseteq\{1,\dots,n\}$ means only
than a researcher observes that $\mathcal{I}_{1}^{\text{treat}}=\{751,752,\dots,1000\}$.
Within Eq (\ref{eq:att-same}), conditioning on $\mathcal{I}_{1}^{\text{treat}}$
is necessary due to the fact that the expectation of some functions
$f\in\mathcal{F}$ may depend on the number of observations used as
input ($n-|\mathcal{I}_{1}^{\text{treat}}|$). Such a dependency holds,
for example, if $f$ involves shrinkage.

One implication of Eq (\ref{eq:att-same}) is that, given a particular
way of partitioning the dataset, if each estimator in $\mathcal{F}$
is unbiased for $E(\tilde{Y}_{i}^{\text{control}}|A_{i}=1)$ then
$\hat{\Delta}_{\text{ATT},F_{\text{split}}}$ is unbiased for the
ATT. Such a property is not true of the blinded procedures discussed
above. As \citet{king2019propensity_not_used} point out, even if
$d(\mathbf{Y})$ is unbiased for all $d\in\mathcal{D}$, the estimator
$\hat{\Delta}_{\text{blind}}=D^{\star}(\mathbf{Y})$ will generally
not be. The key difference is that the data used to select $F_{\text{split}}$
(Step \ref{enu:(Explore)-Freely-explore}) is independent of the data
that is plugged into $F_{\text{split}}\left(\{Y_{i},\mathbf{X}_{i\bcd},A_{i}\}_{i\not\in\mathcal{I}_{1}^{\text{treat}}}\right)$,
while the data used to select $D^{\star}$ is \emph{not} independent
of the data that is plugged into $D^{\star}(\mathbf{Y})$.

Although Algorithm \ref{alg:Sample-splitting-for} can prevent bias,
it does not completely undo all effects of exploration. For example,
p-values produced by Step \ref{enu:(Fit)-Apply-this} may be miscalibrated,
as these values depend on an estimate's entire sampling distribution,
not simply its expectation. The sampling distribution of Algorithm
\ref{alg:Sample-splitting-for}'s output, conditional on a particular
choice of $f$, is not necessarily equivalent to the distribution
that would result from choosing $f$ a priori. Thus, Algorithm \ref{alg:Sample-splitting-for}
is not strictly superior to the traditional sample splitting methods
described above, but rather has different benefits and drawbacks.

Several open research questions remain in terms of how to implement
procedures such as Algorithm \ref{alg:Sample-splitting-for} efficiently,
and how to account for exploration more deeply. For example, consider
the case where $\mathcal{D}=\{d_{\text{OLS},1},\dots,d_{\text{OLS},p}\}$,
and the analyst must choose a single feature to adjust for. An alternative
approach could involve allowing data exploration (either blinded or
unblinded), but also requiring a subsequent test for overfitting based
on how much the distribution of the selected feature changes when
moving to a holdout dataset of covariates. Large changes would indicate
overfitting and potential bias. Approaches in this vein would have
the benefit of not requiring additional outcome data, but also the
drawback of not offering a way to fix the bias that they identify.
Procedures for teams of analysts could be also explored. For example,
each team member could be allowed to explore a different partition
of the data, in isolation, before validating results on the partitions
explored by their peers. Developing such procedures and studying their
formal properties could be a fruitful direction for future research.

\section{Discussion\label{sec:Discussion:-so-what?}}

Rigorous standards of evidence and objectivity are crucial for the
regulatory and litigation settings that motivate outcome blinding.
Unfortunately, and contrary to conventional guidance, we have shown
that the bias incurred by blinded analysts can be on the same order
of magnitude as the bias incurred by unblinded analysts. We have also
outlined a simple, alternative, unblinded procedure for avoiding such
forms of bias, which does not require any outcome data to be discarded
in the final analysis.

An important caveat is that bias from blinded analysts appears to
require that analysts either be explicitly malicious, or have tendencies
towards self-interest. Bias in unblinded scenarios however might plausibly
be the result of well-intentioned analysts second guessing their methods
after seeing a surprising result. For this reason, the insufficiencies
of outcome blinding are most concerning when analysts present reports
to external decision makers with conflicting incentives (e.g., journal
editors, regulatory agencies, or judges). Here, the external decision
makers may wish to enforce safeguards that protect against bias regardless
of its cause. On the other hand, outcome blinding may still be a useful
tool for an analyst reporting to internal decision makers, as the
analyst's team will bear the cost of any poorly informed decision,
and there is less incentive to mislead.

One significant remaining problem manifests under repeated analyses
of shared data. Many observational, retrospective data analyses in
the literature are conducted on \emph{common} datasets that are either
publicly available, or available for a fee. Examples range from modern
databases such as the UKbiobank (\citealp{sudlow2015uk}) and the
SEER Databases on Cancer Statistics (\href{https://seer.cancer.gov/}{www.seer.cancer.gov}),
to the 1987 National Medical Expenditure Survey used by \citeauthor{rubin2001using}
\citeyearpar{rubin2001using,rubin2007design}, which \citeauthor{rubin2001using}
describes as being reused across many litigation cases. For such datasets,
partial information about \emph{in-sample} data is already in the
public domain via published articles. Even if a researcher fully
prespecifies their analysis, it may not be realistic to assume that
the specified analysis was not influenced by previously published
analyses of the same dataset by different labs. This problem is sometimes
referred to as the challenge of developing \emph{quality preserving
data} (\citealp{Aharoni2011-wa,Aharoni2014-lv,Woodworth2018-ri};
see also \citealp{Dwork2015-dm}). Exploring whether methods for controlled
reuse of datasets can be meaningfully combined with some form of blinding
or data partitioning is an important area of future work.

While outcome blinding has become a dominant take-away from \citet{rubin2001using,rubin2007design,rubin2008objective},
and is now somewhat widespread \citep{steiner2010importance,shadish2010primer,yue2012regulatory,yue2014designing,li2016note,ding2017principal,kainz2017improving,lu2019good,king2019propensity_not_used,Chen2021-wl,Levenson2021-ae,ho2021current},
it was never meant to stand in isolation. \citet{rubin2008objective}
also argued that, in addition to blinding outcomes, analysts should
appeal to domain experts when identifying confounders to adjust for.
This advice is vital, since identifying confounders fundamentally
requires knowledge of the underlying causal pathways \citep{pearl2012class_bias_amplifying,vanderweele2019principles}.
Unfortunately, we have also seen that domain expertise can \emph{contribute}
to the bias of blinded, ill-intentioned analysts who choose to use
that expertise to serve their own interests. In this way, involving
domain experts in the analysis does prevent outcome blinding from
being subverted.

Indeed, many modern causal inference approaches do not use blinding,
and instead study the treatment mechanism and the outcome mechanism
simultaneously \citep{de_luna2011covariate_selection,zigler2014uncertainty,shortreed2017outcome,damour2019_deconfounding_score}.
When applied as prespecified procedures, any such method can be studied
through theory and simulation, and should not be dismissed simply
because it does not relegate outcome analysis to a separate stage.

Our general approach is similar in spirit to other quantitative models
of researcher behavior \citep{coker2018theory,fisher2019all} and
of the publication system that researchers inhabit \citep{miller2016optimizing,miller2019quest,patil2016should}.
We recommend that similar models continue to be examined when establishing
best practices for statistical analyses. For example, agent-based
approaches could be relevant when studying publication bias; publication
races between competing labs; the likelihood of a given result being
the subject of a replication study; how data sharing requirements
may change the results that researchers submit; how prediction model
comparisons are presented (see also \citealp{hand2006classifier});
or how courtroom evidence is presented by competing sides. Such studies
could form a valuable complement to empirical studies of how analysts
are observed to behave (e.g., \citealp{leek2011cooperation,jager2014estimate,fisher2014randomized,silberzahn2018many,auspurg2021has}).

\section*{Acknowledgements}

\if0\blind {I am deeply grateful for many helpful conversations and
draft feedback from Leah Comment, Rajarshi Mukherjee, Virginia Fisher,
Daniel Chapksy, Alex D'Amour, and Quanhong Lei, over the course of
developing this work. } \fi

\if1\blind {BLINDED } \fi

\bibliographystyle{chicago}
\bibliography{ps}

\end{document}